# (INVITED) Magnetic resonance probing of ferroelectricity and magnetism in metal-organic frameworks

Nandita Abhyankar [3,5], Sylvain Bertaina [1*], Maylis Orio [2], Naresh Dalal [3, 4]

1 Aix-Marseille Université, CNRS, IM2NP, Marseille, France

2 Aix Marseille Université, CNRS, iSm2, Marseille, France

3 Department of Chemistry & Biochemistry, Florida State University,, Tallahassee, United States

4 National High Magnetic Field Laboratory (NHMFL), Tallahassee, United States

5 Center for Nanoscale Science and Technology (CNST), National Institute of Standards and Technology (NIST), Gaithersburg, United States

* Corresponding author: sylvain.bertaina@im2np.fr

**Abstract:** We employ electron paramagnetic resonance (EPR) of the spin probe $Mn^{2+}$ to study the paraelectric–ferroelectric transition in DMMnF and $Mn^{2+}$-doped DMZnF, which are considered to be model metal−organic frameworks (MOF) with a Pb-free perovskite architecture. In DMAMnF, we study the variation of the $Mn^{2+}$ EPR line shape and intensity at the X-band (~9.4 GHz) and over 80 to 300 K, and we show the absence of magnetoelectric coupling at the ferroelectric transition. At the antiferromagnetic transition in DMMnF, we detect a magnetoelectric coupling caused by weak ferromagnetism in the AFM phase. In DMZnF, the combination of EPR of the $Mn^{2+}$ probe and DFT show that the crystal field is predominantly determined by the $DMA^+$ cations.

## Introduction:

Multiferroics exhibiting simultaneous magnetic and electric ordering offer the possibility of magnetoelectric coupling, which is the electric control of magnetization or magnetic control of dielectric polarization[1-5]. However, reports of magnetoelectric coupling are rare even for single-phase inorganic materials,[6] which have been extensively researched for several decades. Perovskite-like metal formates with the general formula $ABX_3$ are a class of metal–organic frameworks (MOFs) that have received extensive research interest for their ferroelectric and multiferroic properties.[7–23]

Dimethylammonium zinc formate, $[(CH_3)_2NH_2]Zn(HCOO)_3$ (DMZnF), and dimethylammonium manganese formate $[(CH_3)_2NH_2]Mn(HCOO)_3$ (DMMnF) are good prototypes of perovskite-like metal formates and have been studied in detail. DMZnF and DMMnF exhibit a ferroelectric transition at Tc= 156K and 185K, respectively, which is accompanied by a structural phase transition. Figure 1 shows the crystal structure of DMMF (M = Mn, Zn) in the high-temperature (paraelectric) phase. In the paraelectric phase, the $DMA^+$ cation undergoes hindered rotation, with the nitrogen disordered over three equivalent positions. Electron paramagnetic resonance (EPR) is a tool of choice for probing the dynamics of phase transitions. The EPR spectra of $Mn^{2+}$-doped DMZnF are very sensitive to the crystal field modification induced during the ferro/paraelectric phase transition. In DMMnF, the correlation between $Mn^{2+}$ ions is strong enough to create antiferromagnetic order below $T_N$=8 K. Additionally, magnetic correlation is known to hide some spectral properties even in the paramagnetic phase. For example, the 30 EPR lines (caused by splitting of each of the 5 electron spin lines into 6 hyperfine lines) expected for $Mn^{2+}$ are blurred into a single line due to exchange correlation. Even in this case, EPR is sensitive enough to provide information on the coupling between ferroelectric and magnetic orders.

In this paper, we show how the motion of the DMA$^+$ cation can be directly probed by EPR of DMZnF:Mn$^{2+}$. Further, we show that the ferroelectric transition in DMMnF is not accompanied by magnetoelectric coupling, in contrast to what has been suggested before.[15] Finally, we observe weak ferromagnetic/ferroelectric coupling when DMMnF enters the antiferromagnetic phase.

## Experimental details:

Synthesis: Dimethylammonium manganese formate, DMMnF, was synthesized using a standard solvothermal synthetic procedure, described in previous reports.[7] Briefly, 1 mmol of MnCl$_2$·4H$_2$O was dissolved in a solution consisting of 6 mL of dimethylformamide and 6 mL of deionized water. The solution was heated in a Teflon-lined autoclave at 140 °C for 3 days. The supernatant was decanted and left to crystallize, yielding clear, block single crystals after several days. Crystals of Mn$^{2+}$-doped dimethylammonium zinc formate, DMZnF:Mn$^{2+}$, were synthesized by the same method, using a Zn:Mn mole ratio of 1:400 in the initial reaction mixture.

EPR: EPR experiments were performed using a conventional X-band Bruker EMX spectrometer operating at about 9.4 GHz and between 4.5 K and 300 K. A small amount of DPPH was used as a reference, in order to avoid intensity artifacts due to variation of the Q factor of the cavity. Magnetic field modulation associated with lock-in detection was employed, resulting in the derivative of the signal. All measurements were carried out with the static field H perpendicular to the (012) face of the crystal. All the recorded spectra can be fitted by the derivative of a Lorentzian. Simultaneous fits of the DMMnF signal and DPPH signal yield the intensity, line width, and resonance field with high accuracy, and ensure removal of unwanted effects caused by the cavity.

## Results and discussion:

**1- Origin of the crystal field in DMZnF:Mn$^{2+}$.**

Figure 2 shows the EPR signal of Mn$^{2+}$ in DMZnF for T just above and below the ferroelectric phase transition. The spectrum consists of 6 allowed $\Delta S_z=\pm 1$ ($S_z$=-1/2 to 1/2) transitions and 10 forbidden $\Delta S_z=\pm 1$, $\Delta I_z=\pm 1$ transitions. The other allowed transitions ($S_z=\pm 5/2$ to $\pm 3/2$ and $S_z=\pm 3/2$ to $\pm 1/2$) are not resolved due to a large crystal field strain. Note that rotation of the crystal does not change the lineshape dramatically. Below $T_c$, the EPR spectra become much more complex. The $S_z$=-1/2 to 1/2 transitions are sharper and the $S_z=\pm 5/2$ to $\pm 3/2$ and $S_z=\pm 3/2$ to $\pm 1/2$ transitions can now be resolved. For T>$T_c$, the spectrum is rather simple to simulate using the Easyspin module in Matlab. We obtained an axial crystal field anisotropy with D=250MHz, while a very large distribution of D, $\Delta$D=150 MHz, was needed to simulate the linewidth and global shape of the signal. This value of $\Delta$D cannot be explained by a standard local strain, which is usually a perturbation. To explain this value, it is useful to remember that the Mn$^{2+}$ spin is surrounded by 8 nearest-neighbor DMA$^+$ cations (Figure 1). If the timescale of the measurement is fast enough, the DMA$^+$ are seen frozen in one configuration. In the HT phase, there are 3 equivalent positions for each DMA$^+$. Thus, the 8 DMA$^+$ cations surrounding the Mn$^{2+}$ spin can each occupy 3 independent positions, giving $3^8$ = 6561 possible configurations. In the rigid limit, the EPR measurement provides a simultaneous "picture" of all the configurations.

Using quantum chemical methods based on Density Functional Theory (DFT), we have estimated the zero-field splitting of some of these configurations. It is obviously not possible to test all 6561 configurations but a dozen of them were randomly selected and sampled by DFT calculations. Following a previously reported approach,[23] we used minimal models based on X-ray crystallographic measurements and consisting of one hexaformato Mn(II) complex surrounded by 8 DMA$^+$ cations. Here we assume that the zero-field splitting (ZFS) of the system arises mainly from the first and second coordination spheres of the metal ion, i.e. the formate ligands and the DMA$^+$ ions. The distribution of the ZFS parameter was found to be $\Delta D_{DFT}$~125 MHz, which is in rather good agreement with the experimental estimation. Moreover, this result proves that the crystal field variation is determined by the DMA$^+$ cations. For T<Tc, $\Delta D$=15MHz, which is a much more typical value for diluted systems. In this phase, the DMA$^+$ cations are cooperatively ordered, inducing the ferroelectric phase. Each Mn$^{2+}$ sees the same DMA$^+$ configuration and the distribution of crystal fields is greatly narrowed.

## 2- Absence of magnetoelectric coupling at Tc in DMMnF

In DMMnF, magnetic correlation plays an important role and magnetoelectric coupling could appear in the ferroelectric phase.

Temperature-dependence of the EPR of DMMnF was observed in the range from 300 K down to 80 K, with particular attention paid near the structural phase transition (T$_C$ ~ 185 K). Figure 3 shows a series of spectra near T$_C$, obtained in 10 K intervals from T = 210 K to T = 160 K. The sharp peak is due to the DPPH reference. There is a small but clear jump in the height of the DMMnF signal when T$_C$ is crossed.

The spectra were fitted using the derivative of a Lorentzian, and the line width, resonance field, and normalized intensity were extracted from the fits. The parameter of interest for probing magnetoelectric coupling is the variation of the EPR intensity, which is proportional to the static susceptibility. The susceptibility obtained by EPR (Figure 4) shows no abrupt change when the temperature crosses T$_C$, indicating the absence of a magnetoelectric effect. The jump in the height of the EPR spectra is a consequence of the change in linewidth and can be attributed neither to a magnetoelectric nor a magneto-structural effect.[21] This jump is an indirect effect of the change of the EPR linewidth but has no physical meaning related to magnetoelectric coupling.

**3- Evidence of coupling between ferroelectric and weak ferromagnetic orders**

Finally, we focus on the second transition of DMMnF, which is the antiferromagnetic ordering at $T_N$<8.5K. At high T, the EPR line is observed (Figure 5). When the temperature is decreased and approaches $T_N$, the linewidth increases and should diverge at $T=T_N$. Below this temperature, antiferromagnetic resonance (AFMR) lines appear; and these are shifted from the paramagnetic line position because of the internal field. This is the standard behavior of antiferromagnetic systems. However, here we can see that the line is quite convoluted near $T_N$. At T=10K, the line is Lorentzian with a very good signal-to-noise ratio. However, as the temperature is decreased, the signal quality deteriorates due to the appearance of many small peaks of high intensity. This effect is maximum at $T=T_N$, and decreases before disappearing at 4.5 K. This effect is due to the interference of the electromagnetic wave inside the crystal, which acts like an interferometer. Usually, this effect is observed at very high frequencies and for large crystals. It appears when $\lambda(\varepsilon_r\mu_r)^{1/2}=c/f$, where $\lambda$ is the wavelength, $\varepsilon_r$ and $\mu_r$ are the relative permittivity and permeability, respectively, c the velocity of light and f the frequency. $\mu_r$ is usually negligible but near $T_N$, it diverges due to the presence of weak ferromagnetism (canted antiferromagnetism) in the AFM phase.

## Conclusion:

We have shown how EPR can probe the electric and magnetic properties in MOF. Using ferroelectric DMZnF lightly doped by Mn2+, we have shown the dynamic property of the DMA+ cations and the origin of crystal field distribution in the sample. Using the pure magnetic DMMnF system we have shown the absence of magneto-electric coupling at Tc by presence of coupling in antiferromagnetic phase through the weak ferromagnetism .

Acknowledgements:

We would like to thank the interdisciplinary French EPR network RENARD (CNRS−FR3443)

*[(CH3)2NH2]Zn(HCOO)3: Dielectric, 2D NMR, and Theoretical Studies,"* J. Phys. Chem. C 2017, **121**, 6314–6322

Figure 1: Crystal structure of DMMF (M = Mn, Zn) at T>Tc (paraelectric phase). The metal (Zn or Mn) ions are at the center of the oxygen octahedron (red). The 3 possible positions of DMA+ are represented in blue, green, and yellow. At T>$T_c$, the DMA+ is disordered between these 3 equivalent positions. Protons are omitted for clarity.

Figure 2: EPR signal of DMZnF:$Mn^{2+}$, recorded at 9.4GHz and at T just above (red) and below (blue) the ferroelectric phase transition (Tc=150K). For T>$T_c$, only the -1/2 to 1/2 transitions with $\Delta M_I$ = 0 (allowed transitions) and 1 (forbidden transitions) are seen. The other transitions are too broad to be resolved. Below $T_c$, the linewidths of all the lines decrease. The -1/2 to 1/2 transitions are now sharper and the ±5/2 to ±3/2 and ±3/2 to ±1/2 can be resolved.

Figure 3: Comparison of X-band spectra of a DMMnF single crystal in a range of temperatures above and below $T_C$ (~185 K). The small sharp line in the middle is due to the internal standard DPPH. As such, while this observation is in accord with those of Wang et al.,[15] additional studies after double integration showed no jump in the overall signal intensity.

Figure 4: Inverse of EPR susceptibility (green) and line height (orange) as a function of temperature. The susceptibility is extracted from the fit of EPR lines while the height $\delta A$ is measured directly.

Figure 5: EPR Signals of DMMnF for different temperatures from 10 K to 4.5 K. At low temperature, the 2 lines originate from 2 of the 3 structural twins naturally present, while the third signal is out of the range of field. The modulation field was larger than the sharp peaks and is randomly distributed inside the EPR/AFMR signals, proving the non-magnetic nature of these peaks.